\documentclass{article}
\usepackage[margin=1.5in, top=1in, bottom=1in]{geometry}

\usepackage{float}
\usepackage{amsmath}
\usepackage{amsfonts}
\usepackage{xcolor}
\usepackage{booktabs}
\usepackage{multirow}
\usepackage{array,tabularx}
\usepackage{booktabs}
\usepackage{authblk}
\usepackage{graphicx}
\usepackage{subfig}
\usepackage{algorithm}
\usepackage{algorithmic}
\usepackage{tcolorbox}
\usepackage[numbers]{natbib}
\setcitestyle{square,aysep=t,yysep={,}}
\usepackage{fancyvrb}
\DefineVerbatimEnvironment{verbatim}{Verbatim}{xleftmargin=.18in}

\AtBeginDocument{%
	}

\begin{document}
	
	\title{Advanced simulation paradigm of human behaviour unveils complex financial systemic projection}
	
	\author[1, 2]{Cheng Wang}
	\author[1]{Chuwen Wang}
	\author[1]{Shirong Zeng}
	\author[3]{Jianguo Liu}
	\author[1, 2]{Changjun Jiang}
	\affil[1]{School of Computer Science and Technology, Tongji University}
	\affil[2]{the Digital Economy Research Center, Tongji University}
	\affil[3]{Department of Digital Economics, Shanghai University of Finance and Economics}
	
	\maketitle
	
	\begin{abstract}
		The high-order complexity of human behaviour is likely the root cause of extreme difficulty in financial market projections. We consider that behavioural simulation can unveil systemic dynamics to support analysis. Simulating diverse human groups must account for the behavioural heterogeneity, especially in finance. To address the fidelity of simulated agents, on the basis of agent-based modeling, we propose a new paradigm of behavioural simulation where each agent is supported and driven by a hierarchical knowledge architecture. This architecture, integrating language and professional models, imitates behavioural processes in specific scenarios. Evaluated on futures markets, our simulator achieves a 13.29\% deviation in simulating crisis scenarios whose price increase rate reaches 285.34\%. Under normal conditions, our simulator also exhibits lower mean square error in predicting futures price of specific commodities. This technique bridges non-quantitative information with diverse market behaviour, offering a promising platform to simulate investor behaviour and its impact on market dynamics.
		%149个词
	\end{abstract}
	
	\section{Main}
	
	Human behaviour is influenced by multiple factors including experience, cognition, and emotion so that it is always elusive. When a system contains numerous human behaviour, this diverse behaviour interacts between individuals, making the projection of the entire system challenging at a macro level. The financial system represents a typical example of this phenomenon. In such complex human systems \cite{battiston2016complexity, wen2019complexity, ghent2019complexity}, various interpretations of market news and information from investors manifest in their behaviour \cite{chang2015information, ruggeri2020replicating, lin2020evidence}, potentially introducing systemic risks through complex interactions \cite{shiller1990market, li2023information}, especially under abnormal market conditions. Since many complex human systems lack a direct dynamic representation of how diverse interactions within the system affect its overall state in response to a stimulus, approaching the problem from its origin, human behaviour itself, presents a promising solution for computing new system states. Taking the commodities futures market as an example, we investigate whether human behavioural simulation is a constructive methodology to project the trend of the corresponding system.
	
	Simulation has already been an accepted research tool in finance, not because researchers believe simulation is more appropriate, but the difficulty in concluding a resolvable system model. As market hypotheses, theorems, corollaries, and equations have become increasingly complex, it is hard to model the dynamics of market price \cite{Axtell2022Agent}. Researchers have to resort to numerical methodologies, typically simulation from microsimulation \cite{Orcutt1960simulation} to agent-based modeling (ABM) \cite{Samanidou2007agent}, to complement market theory. ABM simulations structure market dynamics at the microscopic level by implementing diverse behavioural patterns and trading algorithms for agents. Early ABMs contributed insights to classical economic theories, such as clustered volatility \cite{arthur1996asset, brock1997rational}. In addition to predefined rules, an agent's behaviour can be specified by modifying the behavioural modeling equations in standard models \cite{arthur1996asset}, or choosing from a set of predefined behaviour patterns \cite{brock1997rational}. With further advances, ABMs began to match real-world statistical patterns \cite{lux1999scaling} and were widely used for market trend prediction \cite{shavandi2022multi}, bubble research \cite{geanakoplos2012getting}, crash analysis \cite{ge2017endogenous}, and contagion modeling \cite{gai2010contagion}. As ABMs proved instructive to some extent, governments and banks adopted them to predict policy impacts \cite{yeh2010examining, darley2007nasdaq}. Machine learning further advanced ABMs by enabling agents to learn from data, identifying patterns and relationships without predefined rules \cite{raman2019financial}. This allows agents to generate more complex and human-like behaviour that is difficult to summarize, which improves their autonomy. Reinforcement learning then enabled agents to dynamically adjust their behaviour based on real-time feedback from the market environment according to their goals, making them more flexible and adaptive \cite{shavandi2022multi, lussange2021modelling, hirano2023neural}. Although arising from a different premise than our view that behaviour is the origin of analysis challenges, financial market simulation has developed significantly in today's data science era.
	
	However, the aforementioned simulation, where agents lack the level of heterogeneity that would make them sufficiently consistent with human behaviour. It is the heterogeneity that generates unpredictable market dynamics. At the individual level, the heterogeneity manifests as personality traits and corresponding diverse behaviour when presented with the same observation. In financial markets, non-quantitative information constitutes a critical component of agents' overall observation. During periods of market turbulence, these factors often exert greater influence on investor behaviour than quantitative historical transaction data. The decision-making processes of various investor classes are frequently dominated by these non-quantifiable elements, particularly when market conditions deviate from normal patterns. Within existing simulation methods, agent's behaviour primarily relies on rule-based models \cite{brock1997rational}, historical data-driven learning \cite{raman2019financial}, or reinforcement learning for decision optimization \cite{shavandi2022multi, lussange2021modelling, hirano2023neural}. The inability to process non-quantitative information further diminishes heterogeneity in ABM simulations of financial markets. This limitation significantly reduces the capacity of models to accurately represent the diverse decision-making patterns exhibited by market participants. There should be a mechanism, advanced to rigid and rational agent modeling, to take non-quantitative elements and their impact on heterogeneous human behavioural patterns into account. Consequently, simulating the flexible and irrational human behaviour observed in response to diverse information in real-world markets will be available \cite{wang2024behavioural}, and so that further aggregated behavioural simulation can be achieved. The challenge of designing such agents has persisted for a long period \cite{matthews2021evolution}. Recent advances in open-domain generative agents and associated simulation techniques present potential solutions to this challenge.
	
	The fundamental prerequisite for human behavioural simulation in real-world complex systems is the agent's possession of comprehensive knowledge, without which performance deteriorates when confronted with the open-domain input out of the agent's input space \cite{huang2020challenges}. Large language models (LLMs), which have become an effective research tool in understanding, reproducing, and affecting human behaviour \cite{NHB1, NHB2, NHB3, NHB4}, address this prerequisite through their accumulation of commonsense knowledge during pre-training \cite{brown2020language}. By leveraging this advanced functionality, daily life simulation in a small town has been realized \cite{park2023generative}. However, market behavioural generation and simulation present additional requirements: 1) equipping agents with varying levels of financial knowledge \cite{bellofatto2018subjective} and 2) agents' behaviourally consistent irrationality observed in real-world market participants \cite{becker1962irrational}. To simulate varying levels of financial knowledge, direct fine-tuning of LLMs requires differentiated design of corresponding training corpora, which not only demands substantial computational resources but also presents significant challenges in corpus creation. Additionally, LLMs struggle to align quantitative trading behaviour with actual market data. Due to their token-by-token generation nature, LLMs must rely on textual representations of numbers when converting decisions into actions. These numerically textual generations inherently carry probability biases from pre-training textual data instead of quantitative data, which is unacceptable for numerical simulation methods. Supplementing financial knowledge using Retrieval-Augmented Generation (RAG) \cite{lewis2020retrieval} presents significant challenges. The development of knowledge bases for market trading is inherently constrained by the scarcity of high-quality open-source materials. Furthermore, during practical inference processes, retrieving information that strongly correlates with specific securities and market conditions proves difficult. Conventional knowledge supplementing methods of LLMs can not match the requirements in both financial reasoning and action process of agents.
	
\begin{figure}[!t]
		\centering
		\includegraphics[width=0.95\linewidth]{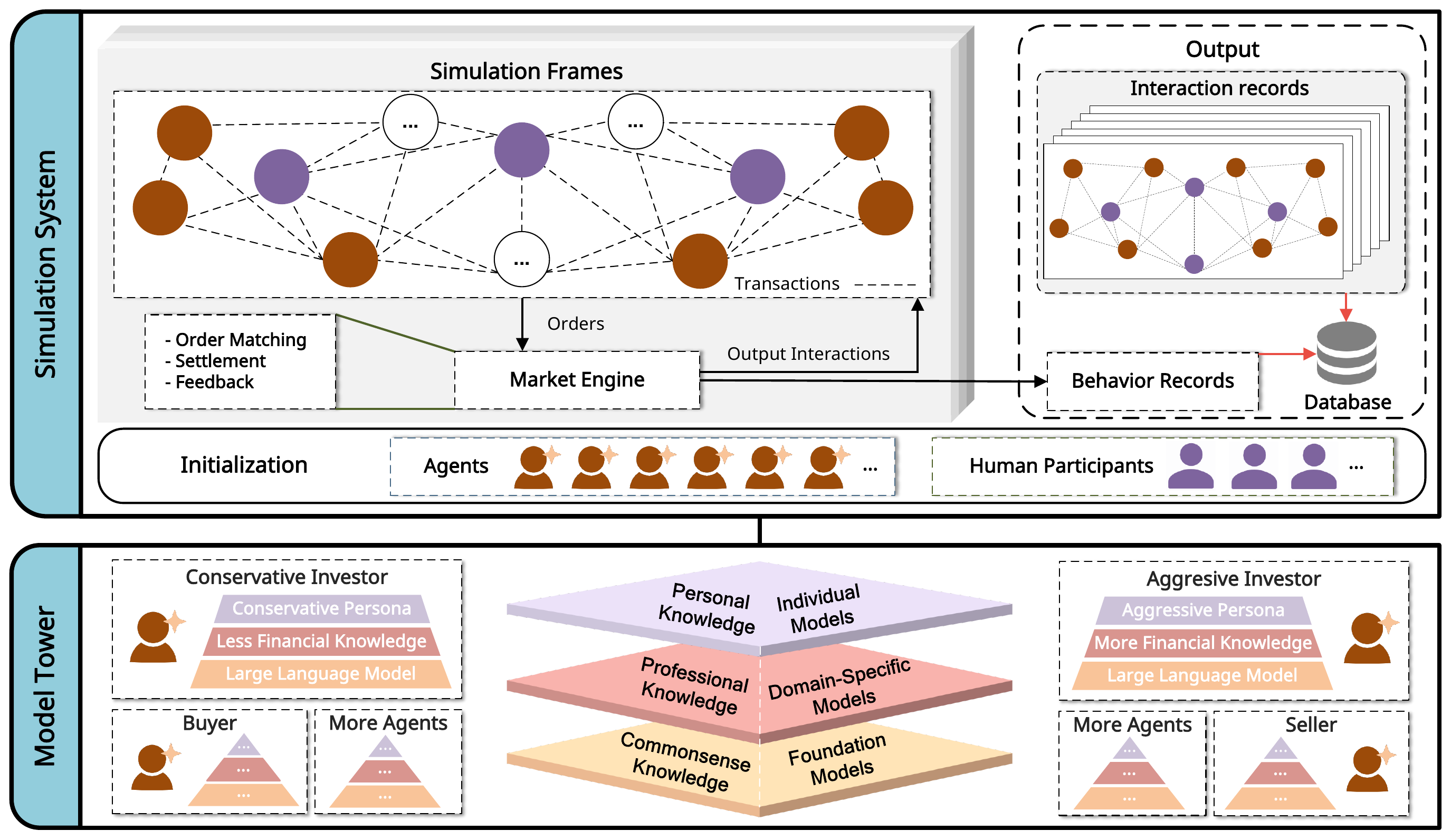}
		\caption{\textbf{Overview of the Market Simulation System.} It organizes the agents to conduct simulation with (or without) human participants chronologically. Each agent is powered by a model tower consisting of hierarchical models.}
		\label{fig:overview}
	\end{figure}

	In this article, we propose a hierarchical knowledge architecture called ``Model Tower'' for generative agents that emulates the knowledge structure of human cognition as displayed in the bottom of Fig.~\ref{fig:overview}. ``Model Tower'' compares hierarchical human knowledge systems to models of different scales and functions. Their mutual complementary and constraining relationship similar to human makes the generated behaviour more aligned with human behaviour. This architecture facilitates the integration of domain-specific financial knowledge and enables the generation of nuanced market behaviour (See the whole architecture in supplementary Note 1). Within the financial market context, we implement a financial expert language model and specialized generator to refine agents' reasoning and actions through financial knowledge integration. Moreover, each agent is endowed with personality and individual knowledge through the system configuration of its LLM. Utilizing these architecturally enhanced generative agents, we develop a financial LLM agent-based market simulation system enabling flexible environment configuration and agent interaction. We evaluate the simulation's effectiveness through a geopolitical futures incident and several futures contracts under normal market conditions, which allows us to examine how the model performs across different scenarios and market environments. The simulation successfully replicates the anomalies induced by the incident and exhibits superior systemic projection accuracy compared to direct applications of LLMs and time-series models in macroeconomic forecasting. We consider that its performance and microscopic interpretability offer valuable analytical and decision-making support.
	
%	\begin{figure}[!t]
%		\centering
%		\includegraphics[width=0.9\linewidth]{pic/ModelTower.pdf}
%		\caption{``Model Tower'' compares hierarchical human knowledge system to models of different scale and functions. Their mutual complementary and constraining relationship similar to human makes the generated behaviours more aligned with human behaviours.}
%		\label{fig:ModelTower}
%	\end{figure}

	\section{Results}
	
	\subsection{Overview of Hierarchical knowledge Architecture}
	
	To enhance agent performance in financial contexts beyond commonsense reasoning, domain-specific financial knowledge integration is essential for accurate market environment interpretation and response generation. We propose a novel framework: financial interactive LLM agents implemented through a hierarchical knowledge architecture comprising commonsense knowledge, professional knowledge, and personal knowledge representations. This architecture is implemented by augmenting the foundation model with domain-specific models that have undergone financial data fine-tuning and personalization through specific cognitive profiles. When processing environmental observations, an agent initially conducts analysis utilizing commonsense reasoning aligned with its configured personal knowledge configurations. To enhance the analytical precision through domain expertise, the agent engages in an iterative consultation process with an expert language model, which generates specialized financial insights. Following multiple iterations of refined reasoning, the action generation phase employs a specialized generator, derived from real-world market transaction data, to transform the agent's textual behavioural tendencies into exact quantitative trading action. This methodology, as opposed to direct quantitative output generation by the LLM, ensures enhanced alignment with real-world investor behaviour patterns while mitigating potential token-based biases inherent in language models during numerical content generation. (More details in Supplementary Note 1.)
	
	\subsection{Simulation of The Tsingshan Nickel Incident}
	
	We choose an influential commodities market incident as the experimental simulation scenario. Compared to stock markets and other financial derivative markets, futures markets feature relatively straightforward participant structures and influencing factors. Under current computational constraints, futures markets represent an appropriate research subject. Their comparative simplicity makes them particularly suitable for modeling studies given the limitations of existing computational resources. At the microscopic level, we expect agents with hierarchical knowledge architecture to respond to a series of market condition changes triggered by sudden geopolitical events in ways consistent with their respective roles. At the macroscopic level, the transactions among agents under market rule calculations should extrapolate abnormal price changes of the underlying assets. If the simulation results at both levels demonstrate credible performance in terms of investor behaviour consistency, error margins in abnormal price predictions, and accurate representation of key event impacts, we consider our agents to achieve, for the first time in market simulation, the capacity to account for non-quantitative information and human factors in market behaviour generation. It also indicates preliminary capabilities to foresee the impacts of sudden abnormal market conditions from the perspective of aggregated human behaviour.
	
	The Tsingshan Nickel Incident of March 2022 was a significant market disruption in the global nickel futures market. It resulted from a combination of geopolitical tensions, supply concerns, and a large short position held by Tsingshan Group, a major Chinese nickel producer. As nickel price on the London Metal Exchange (LME) surged from around 29,000 dollars to over 100,000 dollars per tonne in just two days, LME suspended trading and canceled trades, leading to market chaos and liquidity issues. The incident exemplifies the sudden and unexpected nature of market disruptions, as the geopolitical tensions triggered an unprecedented price spike that no market participant had fully anticipated. This abrupt market shock left traders scrambling to adjust positions with insufficient time to implement proper risk management strategies.
	
	The incident also highlights the strategic game-theoretic aspects of commodity markets, where multiple actors with competing interests engage in complex, multi-layered interactions. Glencore, a major commodities trader, reportedly played a key role in the price surge by aggressively buying nickel contracts, effectively squeezing Tsingshan's short position - a typical example of market participants exploiting vulnerabilities in others' positions. This power struggle between major market players created a strategic game with asymmetric information, where each participant attempted to anticipate and counteract others' moves while protecting their financial interests. The rapid price increase triggered margin calls and forced liquidations, exacerbating the price spike and potentially exposing Tsingshan to billions in losses, demonstrating how strategic decisions by key players can cascade through markets, creating feedback loops that intensify price movements. The event exposed vulnerabilities in the futures market, raised questions about market manipulation and risk management, and had far-reaching consequences for the nickel industry and commodity trading.
	
	\begin{figure}[!t]
		\centering
		\captionsetup[subfloat]{font=small} % 设置子图 caption 字号为 small
		\subfloat[Comparison of price trends]{
			\includegraphics[width=0.31\textwidth]{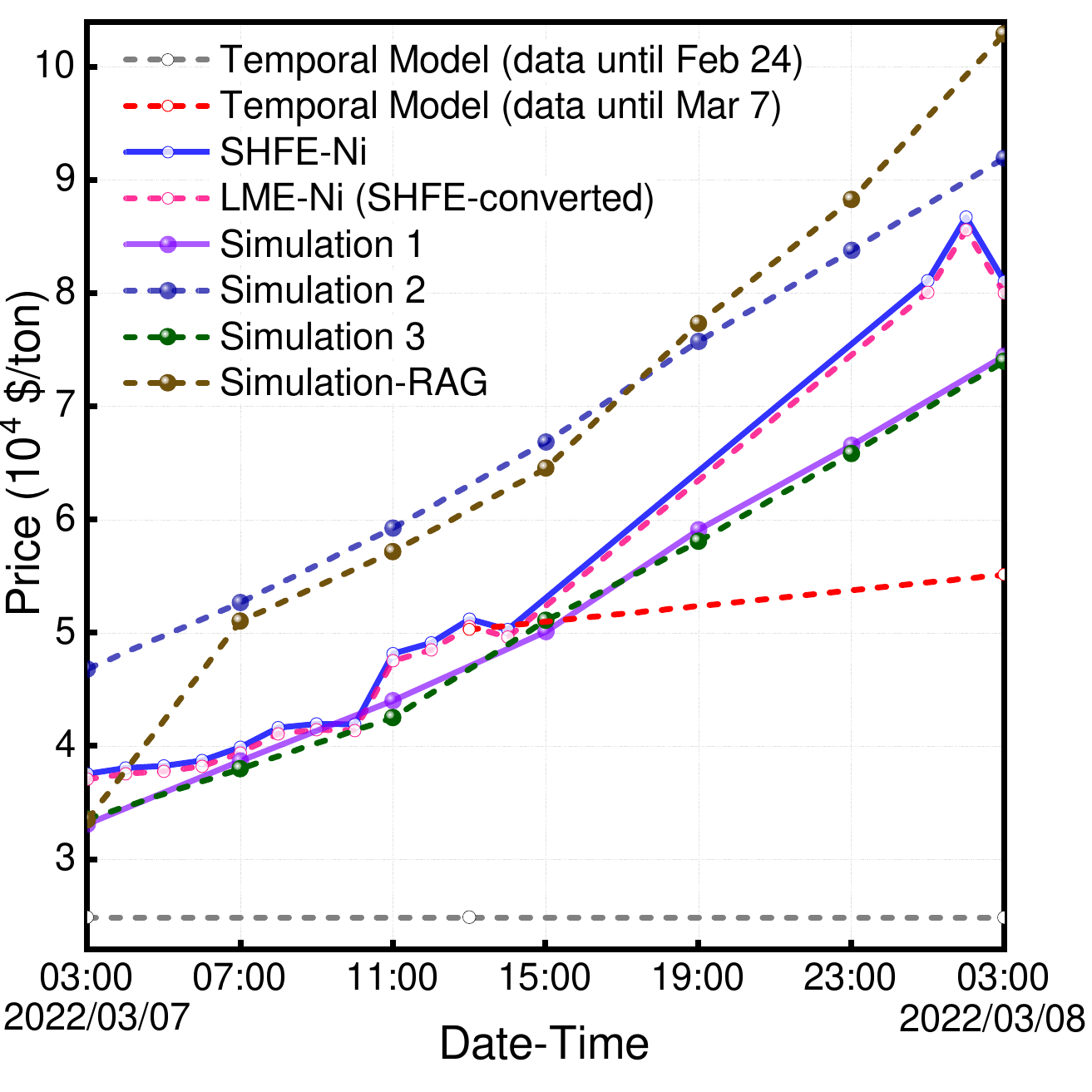}
			\label{fig:trend_comparison}
		}
		\subfloat[Order price (Simulation 1)]{
			\includegraphics[width=0.31\textwidth]{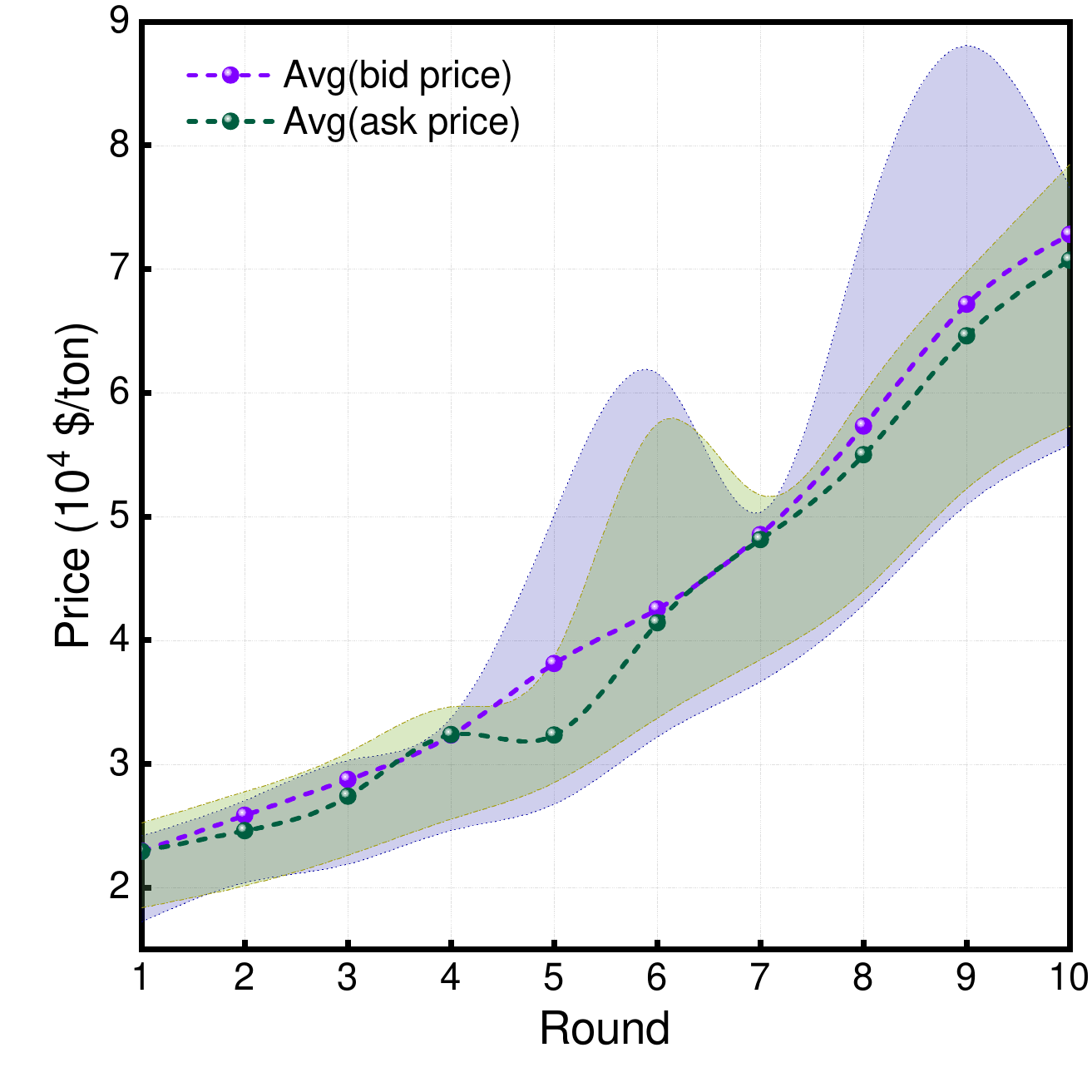}
			\label{fig:order_trend}
		}
		\subfloat[Order price (RAG)]{
			\includegraphics[width=0.31\textwidth]{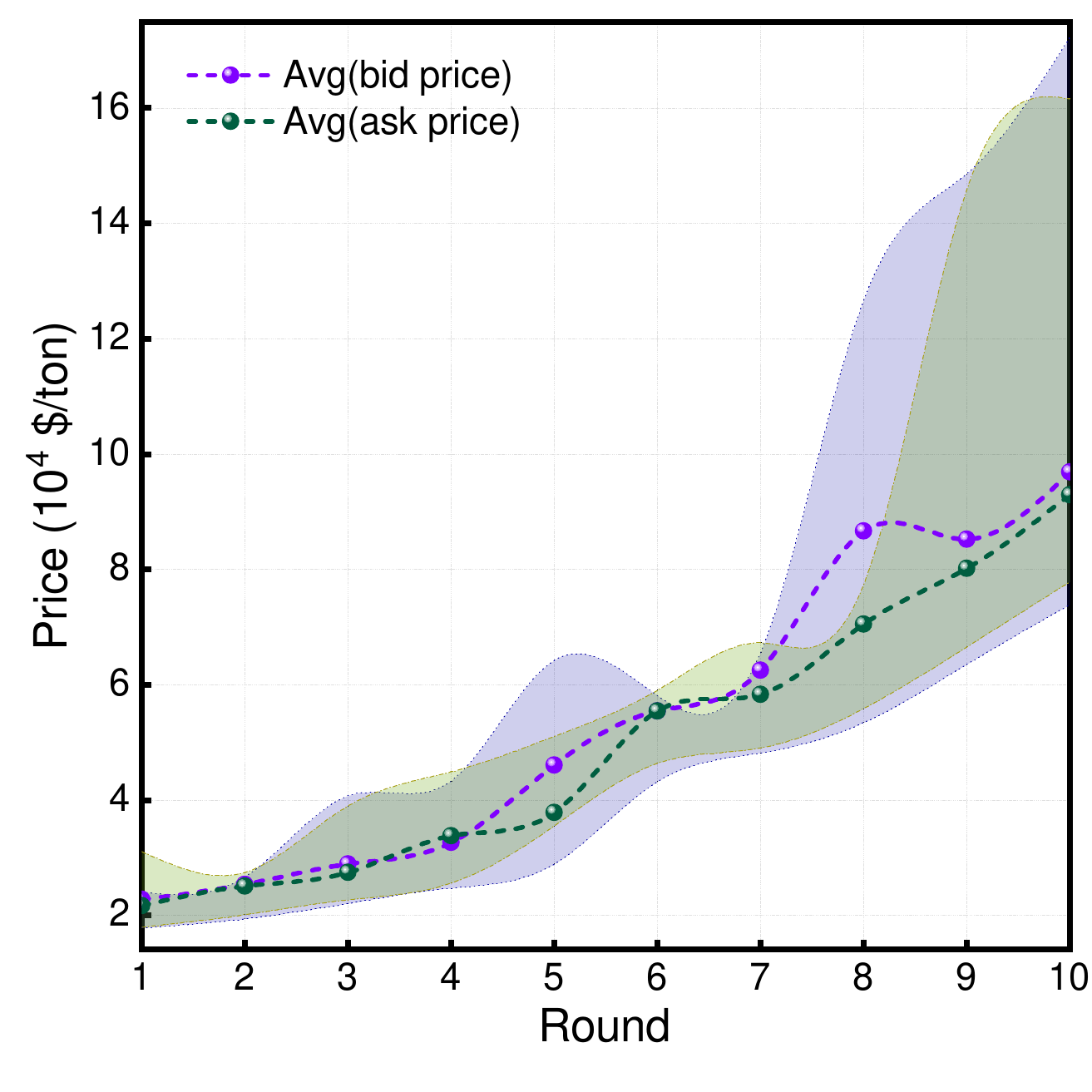}
			\label{fig:price_trend}
		}
		
		\caption{(a) illustrates a comparison among several trajectories of nickel futures price: the prediction from the best-performing general-purpose temporal model (based on data up to February 24th or March 7th, 2022), three simulations generated by our proposed method, a simulation result obtained by replacing the professional knowledge module in our method with a RAG-based approach, and the actual nickel futures price. For the lack of stream price data of LME nickel, its price is converted from the SHFE (Shanghai Futures Exchange) nickel price, which is closely related to the LME nickel price. Glencore began going long on Mar. 7th, 2022, and the market was suspended at 03:10, on Mar. 8th, 2022. Each round of simulated trading represents four hours from round 4 when the Glencore agent begins going long to round 10 when Tsingshan faces large forced liquidation in simulations. (b) and (c) depict the simulated bid and ask price ranges for each trading round, where the vertical spans indicate the highest and lowest price observed. Specifically, (b) corresponds to the simulation result from Simulation 1, while (c) shows the result under the RAG-substituted method using our full model.}
		
		\label{fig:macro}
	\end{figure}
	
	We conduct three simulations with the same initial settings\footnote{Real-world market price data are from Yahoo Finance.} (details in Supplementary Note 2) and an LLM temperature parameter of 1. Fig.~\ref{fig:macro}a illustrates a consistent upward trend in nickel futures price throughout the simulated period, with an accelerated rate of increase following the onset of geopolitical tensions until Tsingshan is unable to meet the margin requirements for all executed contracts due to insufficient funds. It presents a comparative analysis between these three simulation iterations and the actual LME nickel price trajectory. From round 4 to 10, when Glencore begins to go long until the full forced liquidation of Tsingshan Group, the simulated price closely approximates the actual market price. When the actual total price increase reached 285.34\%, the average relative error of the three simulated increases compared to the actual increase was 13.29\%. These macroscopic alignments demonstrate the credibility of our simulated price trajectory. On the contrary, when we substitute the professional knowledge model in our method with a RAG-based approach, the simulated price exhibits significantly more extreme behaviour: the trajectory not only reaches a much higher peak after round 4 when Glencore tries to attack Tsingshan but also displays increased volatility. This deviation can be attributed to RAG's reliance on static retrieval rather than embedded domain priors or temporally structured reasoning. Without the constraint of coherent expert-driven knowledge, the model is more susceptible to amplifying fluctuations in its generative reasoning, leading to exaggerated market responses and unstable price trends. Similarly, the best general temporal model also completely failed to make any prediction about abnormal market conditions, whatever the given data before or after Glencore's beginning to go long. Because the temporal model is neither given nor capable of addressing non-quantitative incident information, it always tends to predict next day's price will fluctuate in a relatively narrow range due to the overwhelming normal market conditions in historical data. Take Simulation 1 as an example (also in the following results), the consistent upward trend of nickel futures price is mirrored in the order price, as evidenced in Fig.~\ref{fig:macro}b. Notably, while the lowest selling price consistently exceeds the lowest bid price in the order book, the highest bid price surpasses the highest ask price in rounds 5, 6, 8, and 9. When compared to the actual price trajectory of LME nickel futures from February 2022 to March 2022 described in the news (pre-modification price is unavailable), our simulation exhibits remarkably similar characteristics. These include a sharp increase preceding the announcement of geopolitical tensions and a subsequent price surge. The bidding behaviour of our agents adheres to established market rules. With the exception of periods dominated by Glencore's long position, the highest bid price consistently exceeds the highest ask price. During a market upward characterized by supply shortages, the average order price of bid price and ask price converge, with buyer price marginally higher. The discrepancy between the simulation using the RAG-based method and the actual price dynamics is also reflected in the order price, as shown in Fig.~\ref{fig:macro}c. The average price shows relatively moderate changes, while the highest bid and ask price exhibits pronounced fluctuations across trading rounds, indicating substantial instability in price generation. Additionally, the gap between the highest and lowest price expands noticeably in later rounds, indicating a widening price range over time.
	
	\begin{figure}[!t]
		\centering
		\captionsetup[subfloat]{font=small} % 设置子图 caption 字号为 small
		\subfloat[Trading behaviour index of agents]{
			\includegraphics[width=0.46\textwidth]{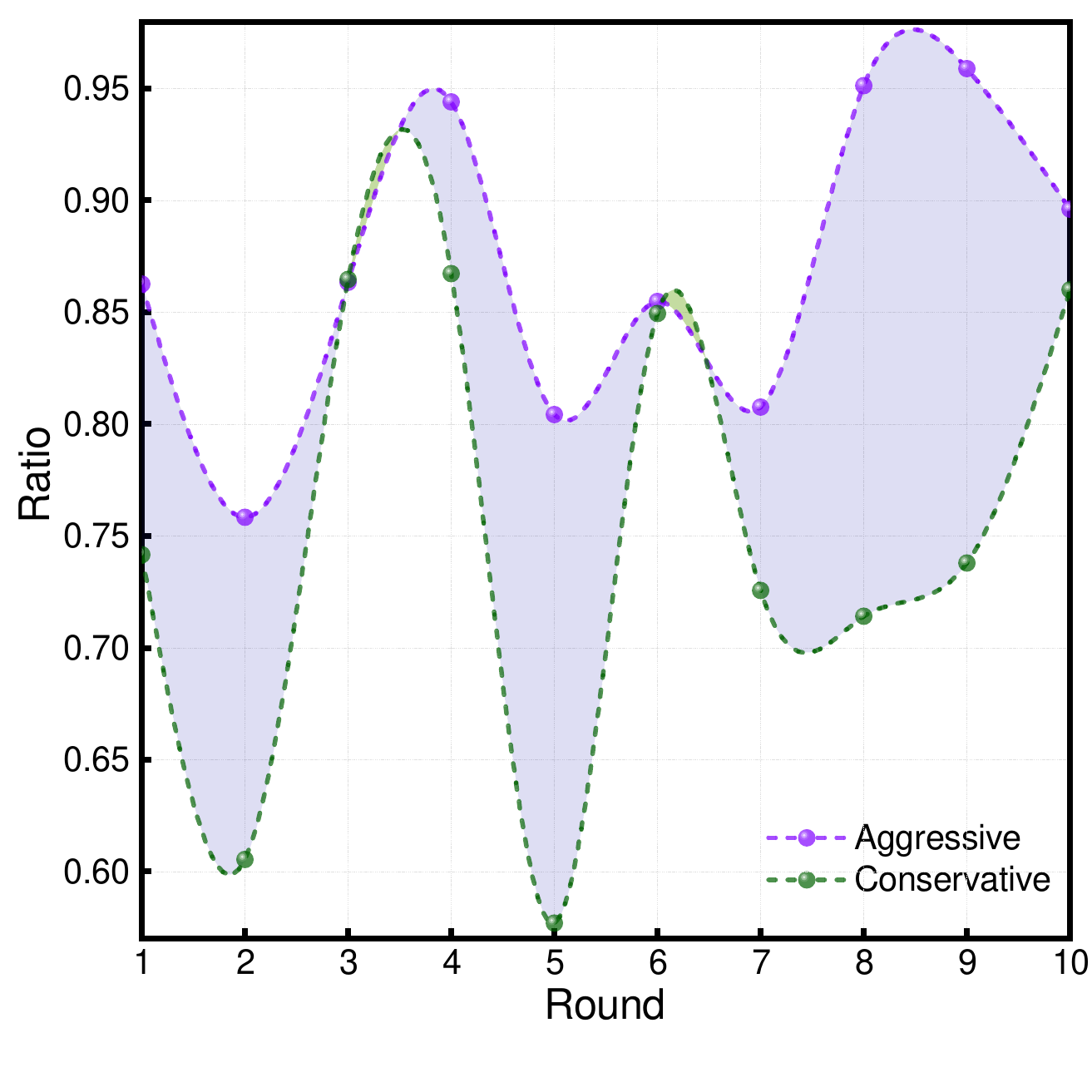}
			\label{fig:agg_con}
		}
		\subfloat[Related order volume]{
			\includegraphics[width=0.46\textwidth]{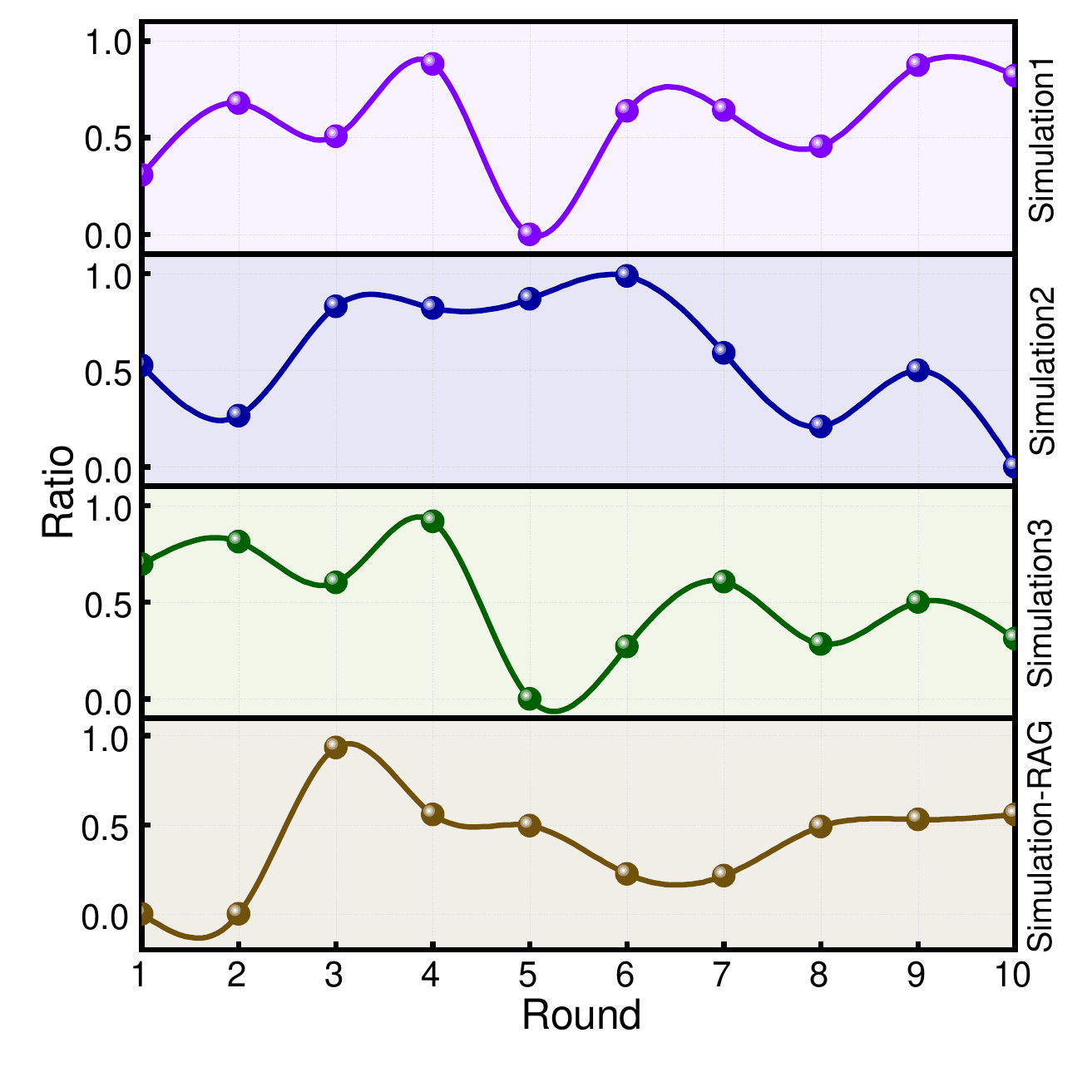}
			\label{fig:buyer}
		}
		\centering
		\quad
		\subfloat[Glencore's completed contracts]{
			\includegraphics[width=0.46\textwidth]{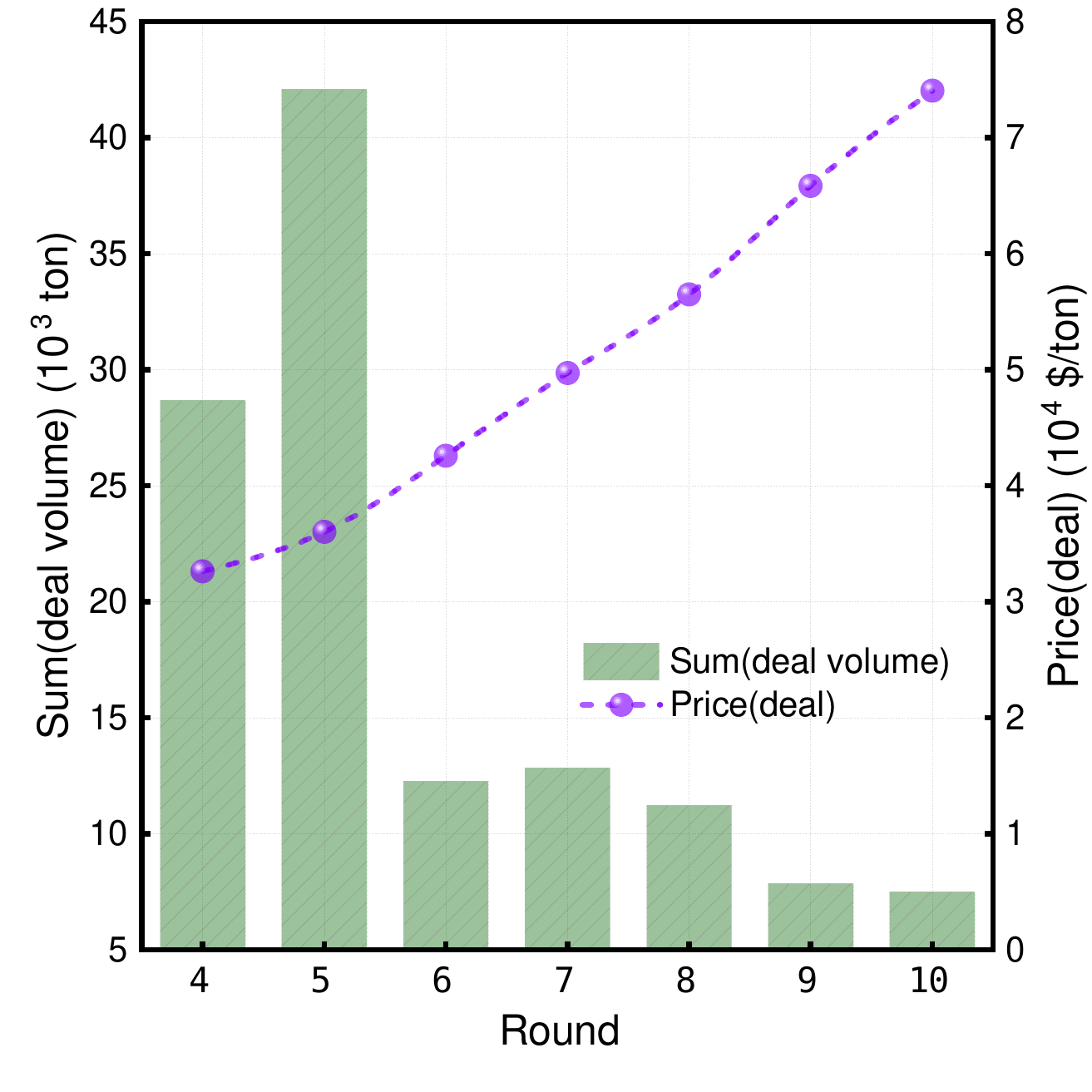}
			\label{fig:GLC_long}
		}
		\subfloat[Cumulative forced liquidation volume]{
			\includegraphics[width=0.46\textwidth]{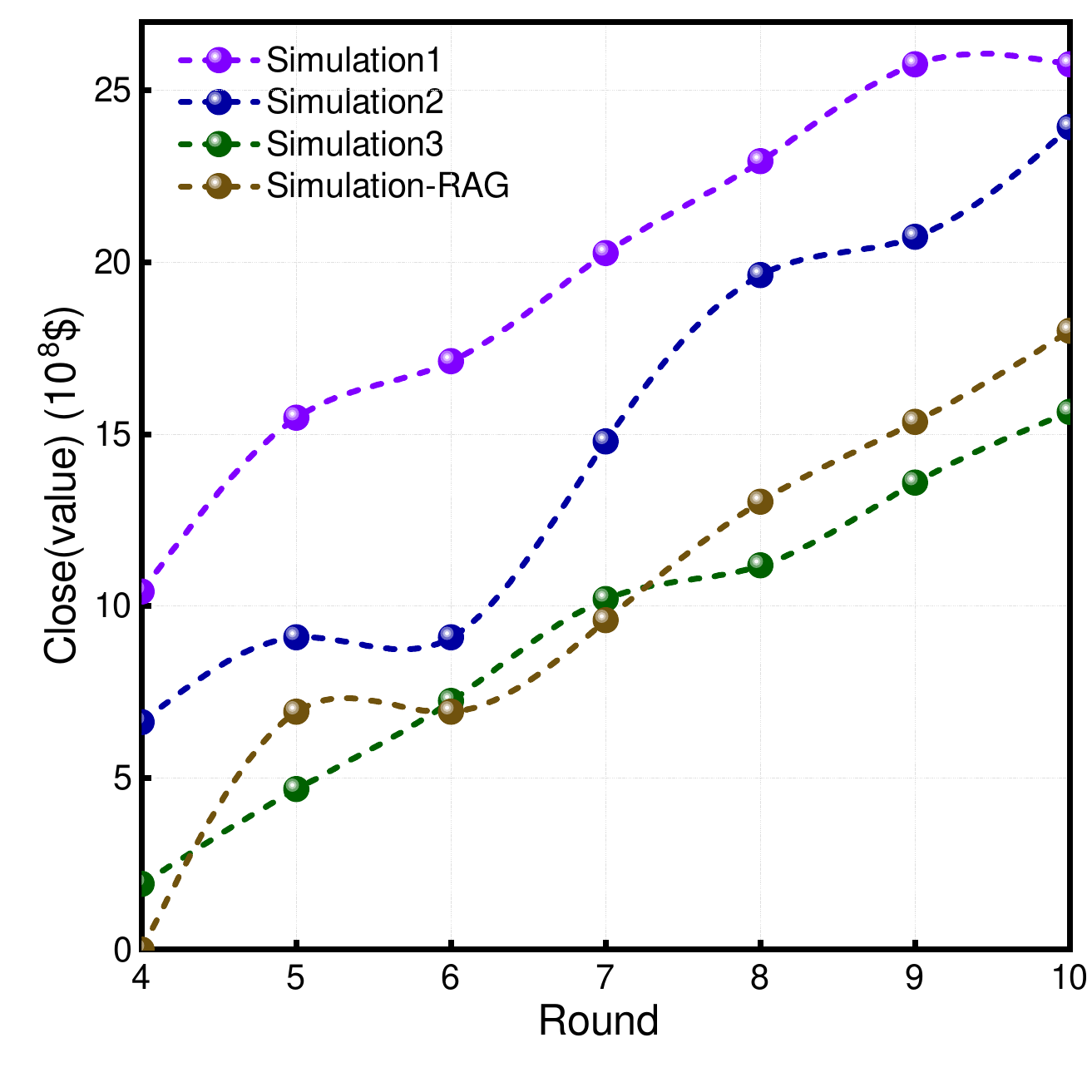}
			\label{fig:TS_liquidation}
		}
		\caption{(a) displays the \textit{trading behaviour index} for each round, representing the ratio of \textbf{actual} futures contracts volume by an agent (conservative or aggressive) at its average order price to the \textbf{maximum} affordable volume, thus characterizing its trading pattern. (b) illustrates the ratio of the buyer agent's buy orders to its maximum affordable volume. (c) presents the volume and weighted-average price of Glencore's completed contracts as nickel buyer in each round from rounds 4 to 10, calculated after each round's settlement phase. (d) illustrates Tsingshan's cumulative forced liquidation volume from rounds 4 to 10, calculated after each round's settlement phase.}
		\label{fig:micro}
	\end{figure}
	
	Fig.~\ref{fig:micro}a reveals a marked disparity in trading behaviour between aggressive and conservative agents. The trading behaviour index of aggressive agents is always greater than that of conservative agents, maintaining an index above 0.75 throughout the simulation. In contrast, conservative agents' index even drops below 0.6 in round 5.
	
	As depicted in Fig.~\ref{fig:micro}b, the buyer agent demonstrates a gradual increase in nickel acquisitions during the initial four rounds. However, a notable shift occurs in round 5, where the agent opts to give up purchases entirely. Besides, the curve from Simulation-RAG diverges significantly from all three other simulations. During the early rounds, the agent under Simulation-RAG places little or no orders, indicating overly conservative behaviour. In the following rounds, although the ratio increases, it remains relatively moderate and fluctuates irregularly.
	
	Aggressive investors consistently allocate a larger proportion of their capital to market transactions compared to conservative investors. In response to influential events that contradict prevailing market trends (as observed in round 5), conservative agents significantly reduce their trading volume. The nickel buyer agent gradually increases its purchase orders in response to the nickel shortage during the initial four rounds. However, in round 5, influenced by news of Tsingshan's intention to short the market, it suspends nickel purchases, anticipating potential price inflation. Following Glencore's long position, the buyer resumes large-scale nickel futures purchases to mitigate raw material costs. These behavioural patterns demonstrate strong alignment between our simulated agents and real-world investors in futures markets.
	
	Fig.~\ref{fig:micro}c illustrates Glencore's trading behaviour, characterized by a significant increase in buy orders starting from round 4, coinciding with the emergence of geopolitical tensions. The executed order volume peaks in round 5, exceeding 40,000 tons, and maintains a level of about 8,000 tons per round thereafter. Concurrently, the average deal price exhibits a rapid ascent, aligning with the overall market price trajectory for nickel futures.
	
	Fig.~\ref{fig:micro}d delineates the growth in Tsingshan Group's cumulative forced liquidation volume. A marked surge is observed following the settlement of round 5, with steady increments in subsequent rounds until round 9. The liquidation volume plateaus at approximately 2.6 billion dollars by the final round. Simulation-RAG also shows a round-by-round increase in forced liquidation, and its overall magnitude is relatively lower than those of the other three simulations. Because it is more difficult for the agent to give orders that are easy to trade without enough financial knowledge.
	
	The results reveal that Glencore capitalizes on the emergence of geopolitical tensions by initiating a substantial volume of buy orders in response to bullish news. When the rumors of Tsingshan's short position circulate in round 5, Glencore anticipates a flood of low-priced sell orders from Tsingshan. Consequently, Glencore places a large number of high-priced buy orders, resulting in a significant volume of completed transactions in this round. Subsequently, Glencore leverages its capital advantage to continue driving up market price through sustained high-priced buy orders. This strategy forces continuous liquidation of Tsingshan's sell orders due to escalating market price until round 9. By this point, all of Tsingshan's previous sell orders are margin-deficient, despite their earlier abandonment of the short position. This simulated sequence of events closely mirrors the real-world forced liquidation process experienced by the Tsingshan group in LME.
	
	\subsection{Ablation Study of Expert Language Model and Specialized Generator}
	
	To evaluate the importance of financial reasoning and action knowledge, we conduct ablation studies on two critical modules of the agent architecture: the expert language model and the specialized generator. In the expert language model ablation, we remove expert advice on market news and trading strategies from the agents. For the specialized generator ablation, we modify the agents' action generation from a tendency-action pattern to a direct action pattern where agents are required to specify precise order quantities and price for input into the market engine, rather than inputting a tendency into the specialized generator. All other system and profile configurations remain consistent with the baseline experiment.
	
	\begin{figure}[!t]
		\centering
		\captionsetup[subfloat]{font=small} % 设置子图 caption 字号为 small
		\subfloat[W/O expert LM]{
			\includegraphics[width=0.31\textwidth]{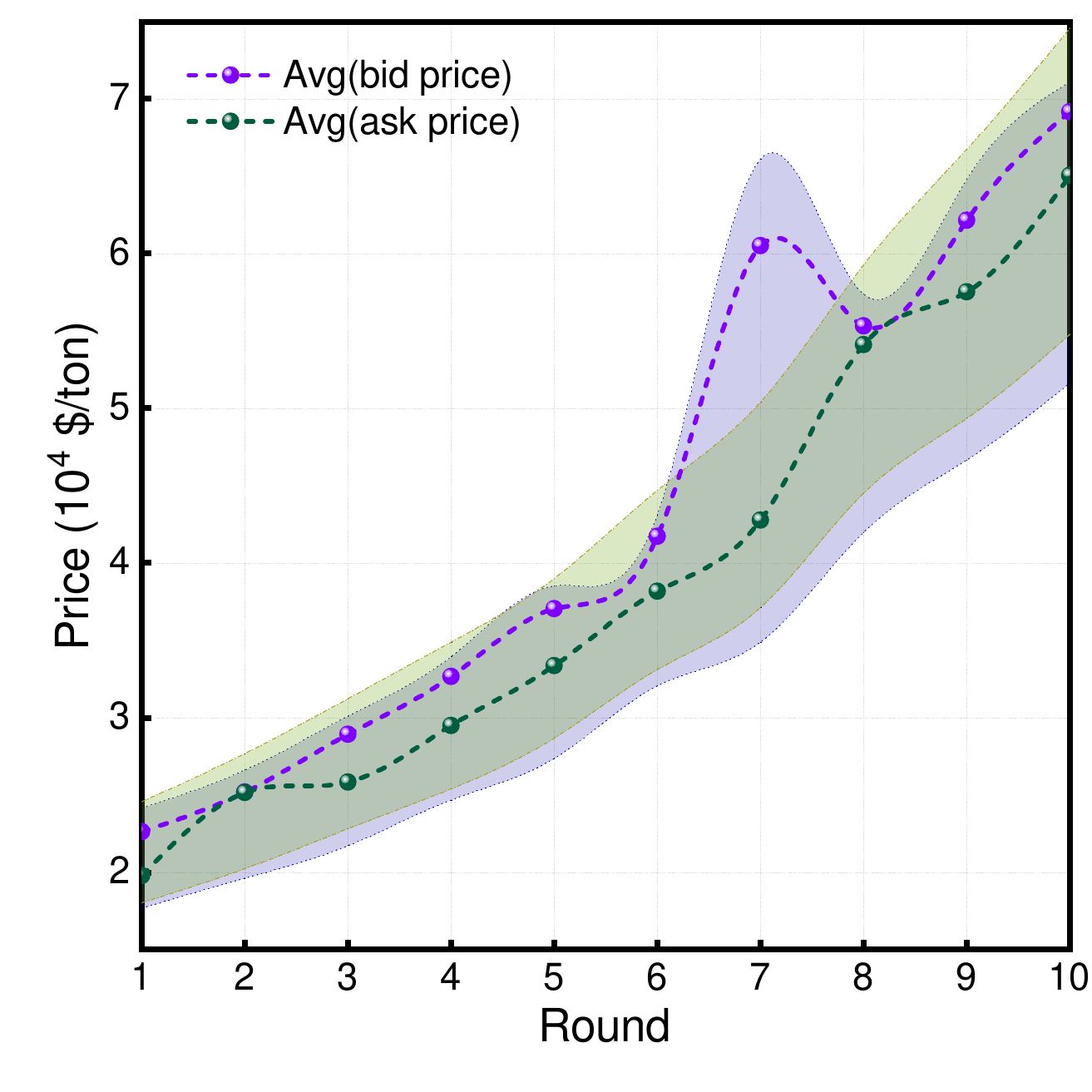}
			\label{fig:ab_1}
		}
		\subfloat[W/O specialized generator]{
			\includegraphics[width=0.31\textwidth]{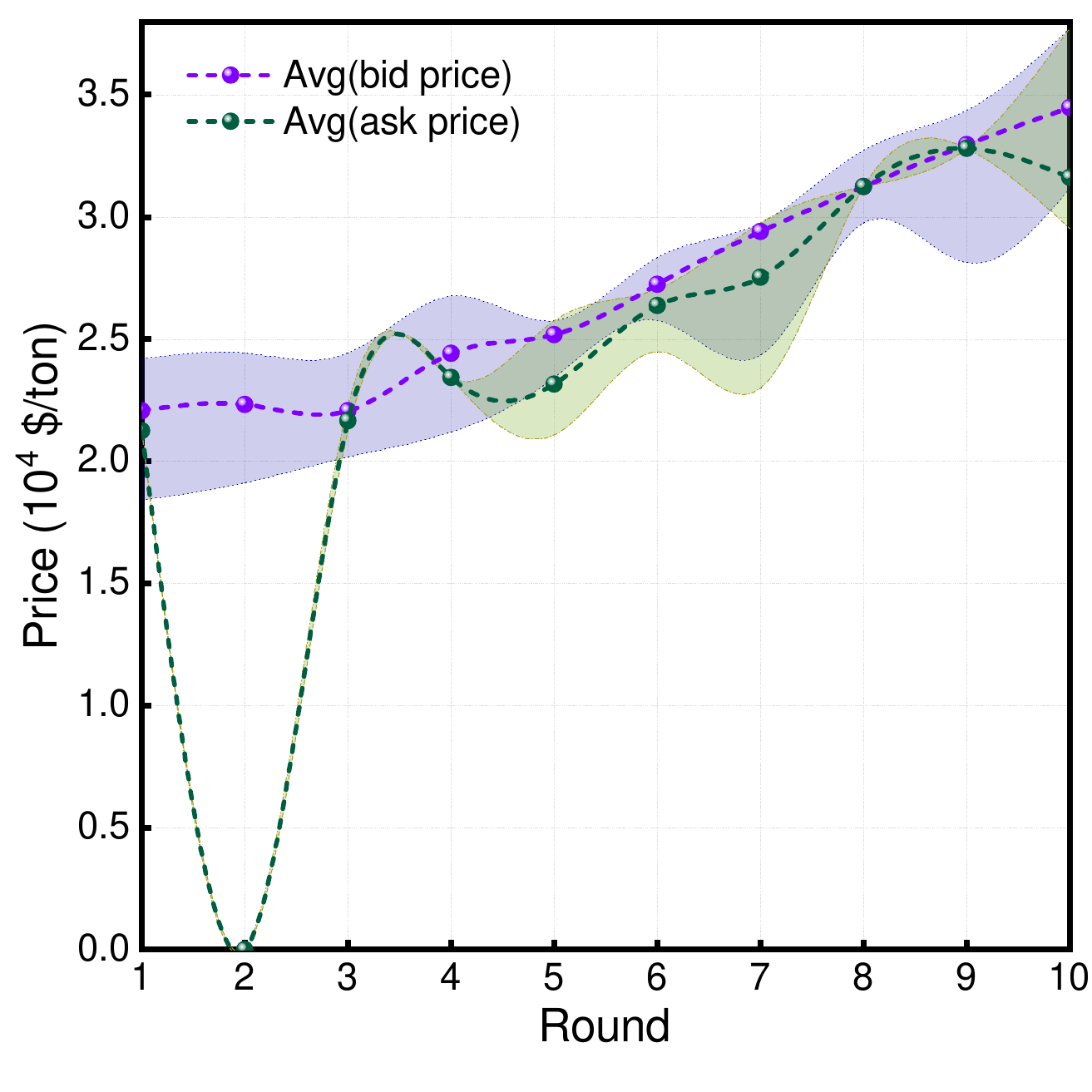}
			\label{fig:ab_2}
		}
		\subfloat[W/O both components]{
			\includegraphics[width=0.31\textwidth]{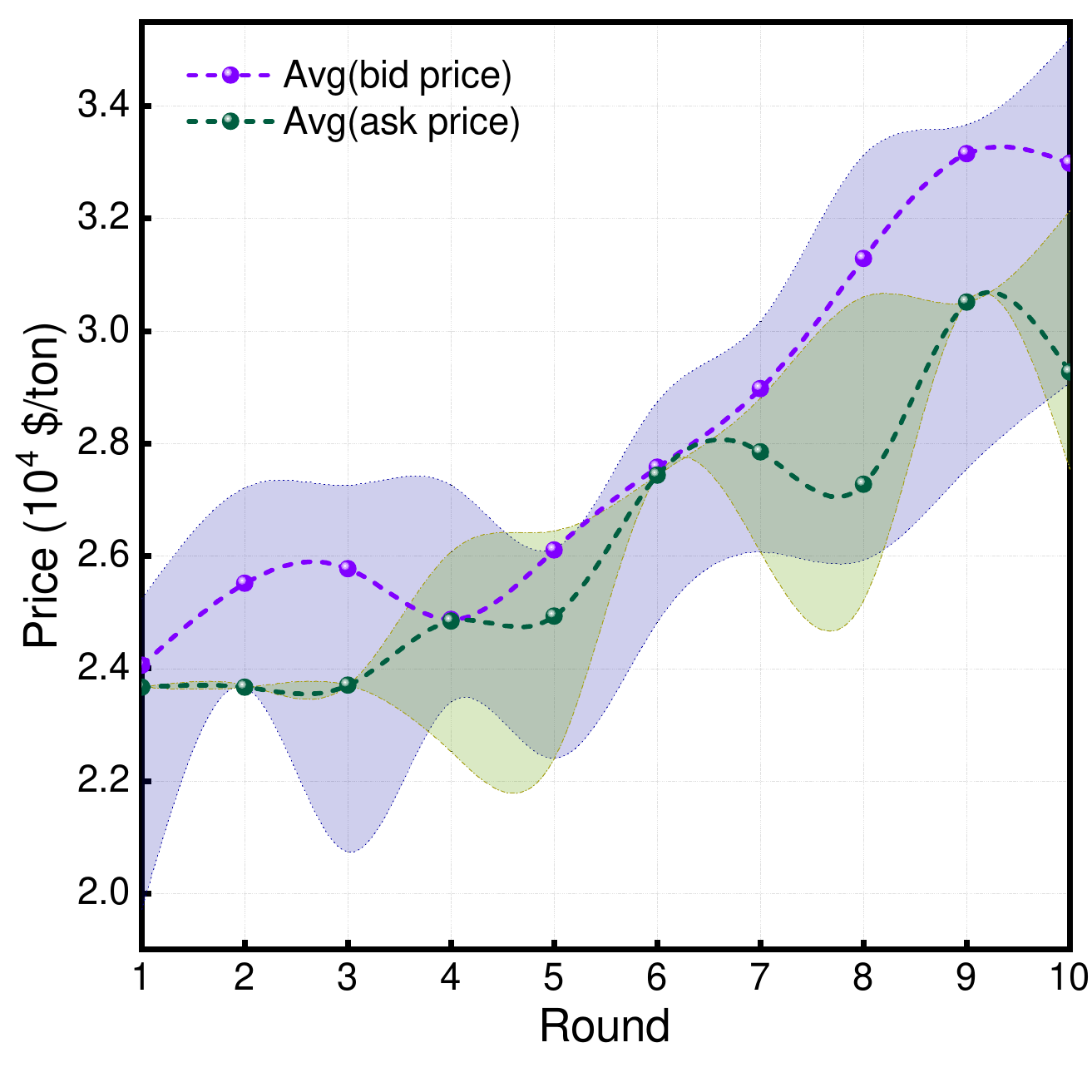}
			\label{fig:ab_3}
		}
		\caption{Price ranges and weighted average price of bid price and ask price across simulation rounds in ablation experiments: (a) Expert Language Model removed, (b) Specialized Generator removed, and (c) Both components removed.}
		\label{fig:ab}
	\end{figure}
	
	Fig.~\ref{fig:ab} presents the results of our ablation studies, illustrating the impact of removing key components from our agent architecture on order pricing dynamics.
	
	In Fig.~\ref{fig:ab}a, we observe the effects of ablating the expert language model. The weighted average price of bid price and ask price demonstrates a consistent upward trajectory across rounds. However, a notable divergence from the baseline simulation is the persistent and substantial gap between the weighted average price of buy and sell orders. 
	
	Fig.~\ref{fig:ab}b depicts the results of the ablation experiment on the specialized generator. In this scenario, the average bid price exhibits relatively stable growth over the rounds. In contrast, the ask price displays volatility and inconsistency. A striking observation is the irregularity in order price ranges, with some rounds showing uniform pricing across all orders. Furthermore, the final round's order price of approximately 35,000 dollars per tonne markedly differs from the baseline simulation's 70,000 dollars per tonne.
	
	Fig.~\ref{fig:ab}c illustrates the outcomes of simulations using only the basic LLM agent, with both the expert language model and specialized generator removed. This configuration results in highly volatile and inconsistent order pricing for both buy and sell orders. Among the various metrics analyzed, only the general price change trend bears resemblance to the baseline simulation. 
	
	Fig.~\ref{fig:macro}c illustrates the outcomes of replacing the Expert Language Model with a RAG module. Compared to the baseline simulation, both the average bid price and its extreme values (highest and lowest) increase significantly,  and the fluctuations become more pronounced with less consistent patterns across simulation rounds. Additionally, the gap between the average and maximum bid price widens noticeably, reflecting greater variability in the agent's bidding behaviour.
	
	The ablation study results reveal that the lack of professional knowledge can lead to the failure in systemic projection. There is a significant disparity in the weighted average price gaps between sell and buy orders when the expert language model is removed from the simulation. Analysis of the agents' inference records indicates that this phenomenon stems from the absence of optimal bidding strategies in trade request orders without expert advice. Similar issues persist when the expert language model is replaced with alternative methods, such as a RAG module, which also fails to provide consistent and context-aware trading strategies. For instance, when faced with excessively high buying price, agents failed to place appropriately higher-priced selling orders in subsequent rounds, which would typically yield greater profits. The results also demonstrate that the absence of the specialized generator leads to a marked reduction and homogenization of the agents' order price ranges. This can be due to the token-by-token generation mode of large language models, where the highest probability candidate word is preferentially selected. When tasked with direct numerical data generation, this characteristic introduces significant bias. In the context of submitted orders, agents instructed to provide specific order price consistently gravitate towards a uniform price point, typically 5\% above the current market price. Consequently, many rounds exhibit minimal order price ranges, with some instances of complete price uniformity. This homogeneity severely compromises the fidelity of market price simulation and undermines the overall rationality of the simulation.
	
	\subsection{Futures Simulation under Normal Market Conditions}
	
	\begin{figure}[!t]
		\centering
		\includegraphics[width=1.0\linewidth]{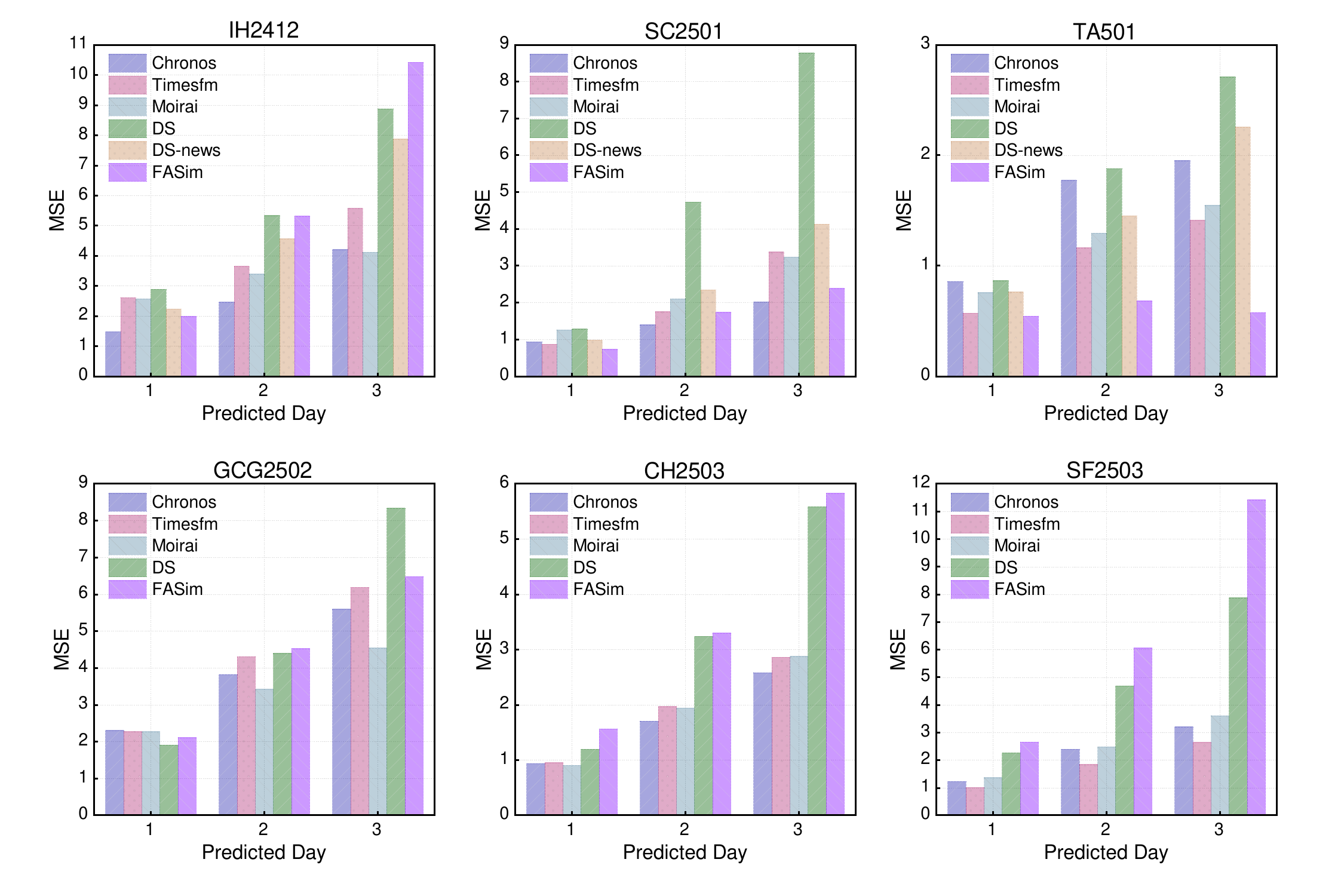}
		\caption{The performance of various prediction methods in forecasting the next three days' settlement price for six futures contracts is evaluated based on the MSE of predicted returns (Chronos: chronos-t5-large, Timesfm: timesfm-2.0-500m, Moirai: morai-1.1-R-large, DS: Deepseek-v2, DS-news: Deepseek-v2 with news, FASim: ours, financial LLM agent-based simulation). For the first three futures contracts, additional news information was incorporated into the predictions using LLMs, i.e. Deepseek and our proposed method.}
		\label{fig:priceforecasting}
	\end{figure}
	
	In addition to reproducing historical events and assisting in risk management, human behavioural simulation exhibits robust generalization capabilities. Specifically, it can simulate trading scenarios for targeted futures contracts by leveraging external information, with the simulated settlement price for each trading day serving as a predictive estimate of the actual settlement price in real-world markets. To evaluate the generalization performance of our simulation method in price prediction, we conduct experiments on six distinct futures contracts. (More details in Supplementary Note 3.)
	
	To evaluate the performance of our behavioural simulation, we compare it against three widely recognized open-source time-series models: chronos-t5-large \cite{ansari2024chronos}, timesfm-2.0-500m \cite{dasdecoder}, and morai-1.1-R-large \cite{woo2024unified}. These models predict future settlement price based on historical settlement price (past 128 trading days) of the futures contracts. Additionally, we include DeepSeek, the LLM integrated into our simulation framework. DeepSeek predicts future settlement price using settlement price from the five days preceding the prediction date, along with monthly and weekly price return rate\footnote{Price data are retrieved from WIND database.} and news data from the current and previous days (when available\footnote{News data in Chinese are from \textit{https://www.eastmoney.com/}.}).
	
	We employ the Mean Squared Error (MSE) of the return rate as the evaluation metric. Let $s_0$ represent the futures settlement price on the last day of historical data, and let $s_i, i\in[1,n]$ denote the actual settlement price of the futures contract on day $i$ in the future, where n is the number of prediction days. The true return rate is $y_i=\frac{s_i-s_0}{s_0}$. Similarly, let $\hat{s}_i, i\in[1,n]$ denote the predicted settlement price on day $i$. The predicted return rate is $\hat{y}_i=\frac{\hat{s}_i-s_0}{s_0}$. The return rate MSE is defined as: $L = \frac{1}{n} \sum_{i=1}^{n} (y_i - \hat{y}_i)^2$. In our experiments, $n=3$.
	
	The experimental results are shown in Fig.~\ref{fig:priceforecasting}. For SC2501, TA501, and IH2412, news information is incorporated into the simulation. For SC2501 and TA501, our method consistently outperforms the baseline models over a three-day horizon, indicating that the simulation framework effectively captures the market impact of news and translates it into more accurate price predictions. However, for IH2412, even with the inclusion of news information, the performance of our method is suboptimal, ranking nearly the lowest among the compared approaches. We hypothesize that this discrepancy arises from the complex market dynamics of stock index futures, which are influenced not only by external information but also by policy changes and macroeconomic conditions—factors that are not adequately expressed in limited news information.
	
	In scenarios where news information is unavailable (GCG2502, CH2503, SF2503), our method performs comparably poorly relative to the baseline models. This suggests that our model relies heavily on news data to predict market fluctuations and make informed decisions. Without such information, the model struggles to capture market volatility, underscoring the critical role of news data in our simulation framework. Furthermore, we observe that DeepSeek's predictive performance is significantly degraded in the absence of news information, highlighting its dependence on external data for accurate market modeling. However, when news information is included, DeepSeek demonstrates substantial performance improvements, emphasizing the importance of external data in enhancing its market understanding and predictive capabilities.
	
	The results highlight both the strengths and limitations of our behavioural simulation in financial markets. It effectively leverages external information, such as news and market price, to generate accurate price predictions for certain futures contracts, particularly in sectors like energy and chemicals. However, its performance deteriorates significantly in the absence of news information or in markets with more complex dynamics, such as stock index futures. These findings suggest that our method is not universally applicable and requires further refinement to improve its adaptability and robustness across diverse market scenarios.
	
	\section{Discussion}
	
	The fundamental challenge in financial market projections stems from the high-order complexity of human behaviour. Our research demonstrates that by simulating heterogeneous group behaviour and analyzing their collective impacts on system states, constructive complex financial systemic projections become achievable. Our experimental findings reveal that this behavioural simulation approach successfully deals with the market impact of non-quantitative factors, providing new insights into complex system dynamics that conventional models cannot capture.
	
	Specifically, this work realizes such simulation by utilizing ``model tower'', and researching the feasibility and value of projecting financial systems from the perspective of human behaviour. As for futures market system, within three simulations of the Tsingshan Nickel Incident, the agents successfully reproduce Glencore Group's long positions leading to market sentiment changes and subsequently Tsingshan Group's substantial forced liquidation. In terms of prediction accuracy, the related error between actual return rate (283\%) and simulated ones is 13\%, while the best general temporal model has a relative error of 99\% (Fig.~\ref{fig:trend_comparison}). Agents representing different populations exhibit behavioural patterns consistent with their characteristics throughout the market simulation process. With agents' better behavioural fidelity, our simulation succeeds in projecting abnormal market Conditions for the first time that none of ABM without the ability to process non-quantitative information can realize. (Our prospect theory experiments in Supplementary Note 4, 5, 6 also demonstrate the fidelity.)
	
	Agents with knowledge systems realize more realistic behavioural simulation. From the simulation results of the Tsingshan Nickel Incident, agents equipped with knowledge systems possess the ability to use comprehensive knowledge to reason about unexpected market events, and to transform these reasoning outcomes into decisions and specific actions. Throughout both decision-making and behaviour generation processes, the diverse financial and personal knowledge bases among agents not only foster more reasonable trading behaviour but also introduce the heterogeneity essential for simulations. According to our ablation experiments, specialized financial knowledge enables agents to consciously and proactively adjust bidding strategies based on market supply-demand relationships and transaction conditions. This pattern of market-driven self-regulation of human behaviour is effectively manifested in the simulation process, which benefits from the professional knowledge guiding agents to make decisions consistent with market participants' common behavioural patterns. In the simulation experiments of normal market conditions with varying volatility levels, the data-based specialized generator learns trader behavioural patterns from transaction data, thereby maintaining behavioural consistency between agents and actual market traders. The comparative results of these experiments demonstrate that agents can, when fully informed, collectively achieve realistic simulations. Such simulations exceed both LLMs' direct predictions and time series models in terms of projection accuracy.
	
	We attribute the success of simulations partly to agent's constraining influence that higher-level models in the knowledge system exert on the general commonsense knowledge base. Numerous behavioural finance findings indicate that market participants consistently exhibit bounded rationality, often making irrational judgments about market conditions. Both LLMs and expert language models have acquired extensive knowledge during their training processes. Due to the token-by-token probabilistic generation mechanism of language models, the rationality level of generated behaviour is inevitably influenced by this mechanism. Certain higher-level models within the knowledge system, such as the specialized generator and personal profile described in our work, impose constraints on this language model mechanism. Rather than compromising overall performance, these constraints actually enhance the agent's ability to produce realistic behavioural simulations. We must acknowledge the success of simulation may be partly attributed to the record of the Tsingshan Nickel Incident in language models' pre-training data despite the mask of time and other identification information in our prompt. Nevertheless, we maintain that the effectiveness of financial LLM agent-based simulation demonstrated in this experiment substantiates its potential for application in analyzing human behaviour in commodity markets and projecting market dynamics under abnormal conditions.
	
	The potential utility of behavioural simulation for systemic projection lies in unprecedented incident impact analyses. For example in commodities futures market, our market simulation system is evident in scenarios such as the aftermath of significant geopolitical events \cite{fernandez2008war, hossain2024impact}. Had this system been available to Tsingshan following the onset of geopolitical tensions, it could have provided valuable insights for market risk management, potentially enabling the company to abandon its short position and mitigate losses. More broadly, it represents a powerful tool for informing critical decision-making processes in complex market environments. Moreover, the flexibility of simulation allows projection based on ``what if'' conditions. Agents initialized with various settings, even in conflict with normal cases can help examine the robustness of systems where our understanding remains incomplete. 
	
	While our study highlights the potential of simulation based on agents with hierarchical knowledge architecture, several frontiers demand further exploration. Scalability remains a critical challenge, as exemplified by our experimental setup requiring approximately 22 hours per simulation with 10 agents under our limited computational resources. Recent advances like \cite{pan2024very} have begun to pioneer the solution. What's more, two directions could improve the simulation system as well. First, our theoretical analysis suggests that a deeper deconstruction of agents' knowledge can continuously improve the alignment between simulated behaviour and human behaviour. Knowledge is not immutable, drawing inspiration from reinforcement learning, an approach worth exploring is learning and updating agent knowledge models at various levels through iterative simulations. Second, the evaluation and utilization of simulation results should also be explored. Our work provides an entry point for understanding the infinite states of complex systems through aggregated human behavioural simulation. How to obtain a better understanding of the latent dynamics embedded in infinite states from limited simulations is also a future direction. We believe that, with continued advancements in this domain, the scalability of interactive simulation systems will become increasingly feasible in the near future.
	
	\section{Methods}
	
	\subsection{Hierarchical Knowledge Architecture for Generative Agents}
	
	The primary objective of enhancing generative agents is to better align their behaviour with the decision-making processes of financially knowledgeable humans in financial markets. We categorize the required financial knowledge into two distinct types: reasoning-related knowledge and action-related knowledge. In our proposed hierarchical knowledge architecture, an expert language model and a specialized generator augment the foundation LLM to incorporate these two types of knowledge respectively.
	
	The expert language model is trained or finetuned on domain-specific data, particularly financial analysis texts such as news reports, corporate financial statements, and social media posts. After the agent formulates an initial reasoning based on observation and commonsense, the expert language model evaluates and refines this reasoning by providing corrections or advice. These outputs are then formatted as input for the agent's reasoning refinement process. The degree to which the agent incorporates these insights varies based on its profile, simulating real-world differences in how investors interpret market information and expert analysis. Following the reasoning refinement phase, the agent makes a final assessment of its observation, expressed as a trading tendency. Through this iterative optimization, the final analysis integrates commonsense reasoning and specialized financial knowledge, enhancing the agent's human-like consistency in financial analysis.
	
	The specialized generator is a generative model designed to transform agents' qualitative trading tendencies into quantitative and concrete actions while accounting for real-world action distributions. We posit that the distribution of actions under specific trading tendencies remains relatively stable, while the distribution of tendencies in response to open-domain news and market-specific information is highly dynamic. However, LLMs typically exhibit systematic biases in generating specific actions due to their token-by-token generation pattern \cite{shanahan2023role}. For instance, when instructed to simulate a dice roll, an LLM agent may disproportionately generate the outcomes=`four' at a frequency significantly higher than the expected probability of $\frac{1}{6}$ \cite{wang2024behavioural}. To mitigate such biases, we introduce the specialized generator, similar to existing tool-using functions \cite{gao2023retrieval}, enabling agents to produce more aligned and accurate actions within the simulation environment. The specialized generator is trained on a categorized trading action dataset, in which action of different trading tendencies and investor styles is explicitly separated. Given a trading tendency and investor style as input, the generator outputs corresponding orders, establishing a structured mapping from qualitative tendencies to specific trading actions. Compared to existing generators that map historical price data to the next orders, our specialized generator leverages well-processed market reasoning information provided by the foundation LLM and the expert language model. This design focuses on a simpler mapping with less market dynamics, enhancing the reliability and authenticity of the generated trading actions.
	
	The financial generative LLM agents in our experiment are constructed using the best pre-trained models. The foundation LLM employed is DeepSeek-V2-0628, a Chinese LLM developed by DeepSeek \cite{deepseek-v2}. For specialized financial analysis, we utilize CFGPT2-7B, a Chinese financial expert language model provided by \citet{li2024ra}. Due to the absence of investor-level data, we implement a specialized order generator that transforms trading tendencies into orders based on a normal distribution with corresponding mean and standard deviation extracted from historical futures price data through $k$-means clustering. (More details in Supplementary Note 8.)
	
	\subsection{Market Simulation System}
	
	The financial LLM agent-based market simulation system is structured into three interconnected modules based on its workflow: the initialization module, the simulation module, and the output module. It integrates agents, a simulation engine, and user interfaces. Researchers primarily control the simulation system through a series of natural language interfaces to configure system settings and actively participate in the simulation. Additionally, the system supports a fully autonomous mode, wherein all market participants are agents, enabling continuous simulation without external intervention. Next, we provide a detailed explanation of the simulation procedure in our system. (Notations are defined in Table.~\ref{tab:notations}, and the algorithm is provided in Supplementary Note 7.)
	
	\begin{table}[!t]
		\caption{Notations of simulation procedure}
		\centering
		\begin{tabular}{cc}
			\toprule
			Notation & Description\\
			\midrule
			$\mathcal{M}$ & The dynamic market state, evolving over time.\\
			$\mathcal{G}$ & The information of traded assets input by user.\\
			$\mathcal{R}$ & The trading rules input by user.\\
			$d_\text{sim}$ & The total number of simulation ``frames''.\\
			$d_\text{turn}$ & The number of trading turns per frame.\\
			$\mathcal{E}$ & The simulation engine.\\
			$\mathcal{P}$ & The predefined market rules. \\
			$N$ & The number of agents.\\
			$\mathcal{A}=\{\mathcal{A}_i\}_{i=1}^{N}$ & The set of agents.\\
			$\mathcal{A}_i^\text{profile}$ & The profile of the $i$-th agent.\\
			$\mathcal{A}_i^\text{account}$ & The account information of $\mathcal{A}_i$.\\
			$\mathbf{A}$ & The set of transaction records storing all simulation-generated data.\\
			$\mathcal{M}_t$ & The user-input environmental information at frame $t$.\\
			$m_{i,t}$ & The market assessment of $\mathcal{A}_i$ at frame $t$.\\
			$\mathbf{O}_\text{match}$ & The set of order requests waiting to be matched.\\
			$\mathbf{O}_\text{success}$ & The set of successfully matched order requests.\\
			$\mathbf{O}_\text{withdraw}$ & The set of unmatched order requests withdrawn by agents.\\
			$s_{i,t,k}^\text{init}$ & The preliminary trading strategy formulated by $\mathcal{A}_i$ at turn $k$ of frame $t$.\\
			$s_{i,t,k}^\text{fina}$ & The refined trading strategy by agent $\mathcal{A}_i$ at turn $k$ of frame $t$.\\
			$o_{i,t,k}$ & The order requests generated by $\mathcal{A}_i$ at turn $k$ of frame $t$.\\
			$s_i=\{s_{i,t,k}^\text{fina}\}_{k=1}^{d_\text{turn}}$ & The set of final strategies adopted by $\mathcal{A}_i$ during frame $t$.\\
			$o_i=\{o_{i,t,k}\}_{k=1}^{d_\text{turn}}$ & The set of order requests submitted by $\mathcal{A}_i$ during frame $t$.\\
			$r_{i,t}$ & The reflection generated by $\mathcal{A}_i$ on its past strategies and actions in frame $t$.\\
			\bottomrule
		\end{tabular}
		\label{tab:notations}
	\end{table}
	
	The initialization module provides two primary interfaces: the simulation engine configuration interface and the agent configuration interface. 
	\begin{enumerate}
		\item \textbf{Engine Initialization}: Through the simulation engine configuration interface, researchers specify the market conditions $\mathcal{M}$, asset information $\mathcal{G}$, trading rules $\mathcal{R}$, and simulation duration $(d_\text{sim},d_\text{turn})$, where $d_\text{sim}$ is the total number of simulation rounds, and $d_\text{turn}$ represents the number of trading turns per round. The simulation engine $\mathcal{E}$ is then initialized based on these parameters and predefined market rules $\mathcal{P}$ that emulate real-market mechanisms, such as deal execution and account settlement protocols.
		\item \textbf{Agents Configuration}: The agent configuration interface allows researchers to define the number of agents $N$, their profiles, account information, and model configurations (including model selection, temperature, and top\_p values for our agent implementation). Researchers can also configure an agent as their proxy for active participation in the simulation. The set of agents $\mathcal{A}=\{\mathcal{A}_i\}_{i=1}^{N}$ is then initialized based on these parameters, where each agent $\mathcal{A}_i$ has a profile $\mathcal{A}_i^\text{profile}$ and an account $\mathcal{A}_i^\text{account}$. 
		\item \textbf{Data synchronization}: Once all configurations are set, the engine $\mathcal{E}$ synchronizes the data across agents $\mathcal{A}$ and the database to ensure consistency. An action record set $\mathbf{A}$ is initialized as an empty set to store transaction data.
	\end{enumerate}
	
	The simulation module executes $d_\text{sim}$ iterative simulations in time steps, or ``frames'', with two primary interfaces: the market environment configuration interface and the user market action interface. 
	
	At the beginning of each frame $t(t\in [1,d_\text{sim}])$, new environmental information $\mathcal{M}_t$, such as news or agent-specific information, can be introduced via the market environment configuration interface to influence the simulation dynamics. Subsequently, researchers can either actively participate by submitting order requests (in a format consistent with agent inputs) or passively observe agent interactions. For example, during the trading phase, researchers can enhance simulation realism by directly submitting requests through the user market action interface rather than guiding their designated agent. Alternatively, their designated agents operate autonomously based on predefined settings, enabling fully automated simulation. The market observation available to agents, denoted as $O_t$, consists of $\mathcal{M}_t$ and the latest market data $\mathcal{M}$ (e.g., asset price, major holders' positions) retrieved from the database via engine queries. Each agent $\mathcal{A}_i(i\in [1,N])$ processes $O_t$ to analyze market conditions and assess trends. To enhance its analysis, $\mathcal{A}_i$ consults an expert language model, integrating domain-specific financial insights into its decision-making process. The final market trend assessment is recorded as $m_{i,t}$. Once all agents complete their analysis, the trading phase begins.
	
	The trading phase in frame $t$ comprises $d_\text{turn}$ consecutive turns. At the onset of this phase, the set of order requests $\mathbf{O}_\text{match}$ is initialized as empty. During each turn $k(k\in [1,d_\text{turn}])$, every agent $\mathcal{A}_i$ performs the following steps:
	\begin{enumerate}
		\item \textbf{Account State Retrieval}: $\mathcal{A}_i$ queries the engine $\mathcal{E}$ for its latest account information $\mathcal{A}_{i}^\text{account}$, including total funds, available funds, profit and loss, and asset holdings.
		\item \textbf{Preliminary Trading Strategy Formation}: Based on its account state $\mathcal{A}_{i}^\text{account}$, market analysis $m_{i,t}$, and reflection $r_{i,t-1}$ from the previous frame, the agent formulates a preliminary trading strategy $s_{i,t,k}^\text{init}$.
		\item \textbf{Strategy Refinement via the Expert Model}: The agent refines its strategy $s_{i,t,k}^\text{init}$ using the expert language model to enhance reasoning and decision quality, formulating the final strategy $s_{i,t,k}^\text{fina}$.
		\item \textbf{Order Requests Generation}: The agent utilizes a specialized generator to transform the qualitative trading tendency $s_{i,t,k}^\text{fina}$ into quantitative order requests $o_{i,t,k}$, specifying transaction volume, price, and type (e.g., buy or sell). The generated requests $o_{i,t,k}$ are appended to $\mathbf{O}_\text{match}$.
	\end{enumerate}
	
	After all agents and researchers submit their order requests, the engine $\mathcal{E}$ matches requests from $\mathbf{O}_\text{match}$ according to trading rules $\mathcal{R}$ and generates a set of successful deals $\mathbf{O}_\text{success}$. Matched requests in $\mathbf{O}_\text{success}$ are removed from $\mathbf{O}_\text{match}$, and both $\mathbf{O}_\text{success}$ and $\mathbf{O}_\text{match}$ are recorded in $\mathbf{A}$ and stored in the database. The matching results in turn $k$ are fed back to each agent $\mathcal{A}_i$ by the engine $\mathcal{E}$. Upon receiving this feedback, $\mathcal{A}_i$ determines whether to withdraw unmatched requests. The set of withdrawn requests $\mathbf{O}_\text{withdraw}$ is removed from $\mathbf{O}_\text{match}$.
	
	At the conclusion of the trading phase in frame $t$, the engine $\mathcal{E}$ automatically withdraws all remaining unmatched requests in $\mathbf{O}_\text{match}$. Then it executes settlement for all deals across the $d_\text{turn}$ turns based on $\mathcal{P}$, updating market states $\mathcal{M}$ and individual agent accounts $\mathcal{A}_{i}^\text{account}(i\in[1,N])$ accordingly. At the termination of frame $t$, analogous to the close of a trading day, each agent $\mathcal{A}_i$ conducts reflection of its past strategies $s_{i}=\{s_{i,t,k}^\text{final}\}_{k=1}^{d_\text{turn}}$ and actions $o_{i}=\{o_{i,t,k}\}_{k=1}^{d_\text{turn}}$. This reflection $r_{i,t}$ serves as input for decision-making in the subsequent frame ($t+1$), simulating an adaptive learning process akin to real-world investor behaviour.
	
	The simulation concludes after $d_\text{sim}$ rounds. The output module constructs a database to store all recorded data, preserving agent interactions as graph-structured data frames and storing transaction records in relational tables. These datasets are accessible through an analysis interface, enabling researchers to conduct manual market risk management studies and further evaluation based on the simulation outcomes.

\end{document}